%% file: main.tex
\documentclass[sigconf]{acmart}

\usepackage{booktabs} 

\usepackage[english]{babel}
\usepackage{moresize}
\usepackage{amsmath}
\usepackage{algorithmic} 
\usepackage{balance}
\usepackage{comment}
\usepackage{paralist}
\usepackage{bm}
\usepackage{pgfplots}
\pgfplotsset{compat=1.17}
\usetikzlibrary{pgfplots.dateplot}

\usepackage{flushend}
\usepackage[english]{babel}
\usepackage[latin1]{inputenc}
\usepackage{mathrsfs}
\usepackage{graphicx}

\usepackage{amssymb}
\usepackage{amsfonts}
\usepackage{url}
\usepackage{longtable}
\usepackage{rotating}
\usepackage{multirow}
\usepackage{mathrsfs}
\usepackage{subfigure}
\usepackage{enumitem}
\usepackage[linesnumbered,algoruled,boxed,lined]{algorithm2e}
\usepackage{adjustbox}
\usepackage{hyperref}

\usepackage{pgfplots}
\usetikzlibrary{pgfplots.dateplot}

\usepackage{diagbox}

\usepackage{filecontents}
\definecolor{tblue}{RGB}{31,119,180}
\definecolor{torange}{RGB}{255,127,14}
\definecolor{tgreen}{RGB}{44,160,44}
\definecolor{tred}{RGB}{214,39,40}
\definecolor{tpurple}{RGB}{148,103,189}

\newcommand{\hide}[1]{} 

\newcommand{\etal}{\textit{et al}.}

\newcommand{\ie}{\textit{i}.\textit{e}.}
\newcommand{\eg}{\textit{e}.\textit{g}.} 
\newcommand{\wrt}{\textit{w}.\textit{r}.\textit{t}}

\setcopyright{acmcopyright}

\def\full{Contrastive Meta Learning}
\def\model{CML}
\begin{document}
\fancyhead{}

\title{Contrastive Meta Learning with Behavior \\ Multiplicity for Recommendation}

\author{Wei Wei$^{1,3}$, Chao Huang$^{1,2*}$, Lianghao Xia$^{1}$, Yong Xu$^{3}$, Jiashu Zhao$^{4}$, Dawei Yin$^{5}$}
\thanks{$*$ Corresponding author: Chao Huang}
\thanks{This work was done when Wei was an intern under supervision of Chao Huang}
\affiliation{$^1$Department of Computer Science, $^2$Musketeers Foundation Institute of Data Science, University of Hong Kong \\
$^3$South China University of Technology, $^4$Wilfrid Laurier University, $^5$Baidu Inc \\
weiwei1206cs@gmail.com, chaohuang75@gmail.com, aka\_xia@foxmail.com, \\yxu@scut.edu.cn, jzhao@wlu.ca, yindawei@acm.org\\
}

\begin{abstract}
A well-informed recommendation framework could not only help users identify their interested items, but also benefit the revenue of various online platforms (\eg, e-commerce, social media). Traditional recommendation models usually assume that only a single type of interaction exists between user and item, and fail to model the multiplex user-item relationships from multi-typed user behavior data, such as page view, add-to-favourite and purchase. While some recent studies propose to capture the dependencies across different types of behaviors, two important challenges have been less explored: i) Dealing with the sparse supervision signal under target behaviors (\eg, purchase). ii) Capturing the personalized multi-behavior patterns with customized dependency modeling. To tackle the above challenges, we devise a new model \model, \textbf{\underline{C}}ontrastive \textbf{\underline{M}}eta \textbf{\underline{L}}earning (\model), to maintain dedicated cross-type behavior dependency for different users. In particular, we propose a multi-behavior contrastive learning framework to distill transferable knowledge across different types of behaviors via the constructed contrastive loss. In addition, to capture the diverse multi-behavior patterns, we design a contrastive meta network to encode the customized behavior heterogeneity for different users. Extensive experiments on three real-world datasets indicate that our method consistently outperforms various state-of-the-art recommendation methods. Our empirical studies further suggest that the contrastive meta learning paradigm offers great potential for capturing the behavior multiplicity in recommendation. We release our model implementation at: https://github.com/weiwei1206/CML.git.


\end{abstract}



\copyrightyear{2022}
\acmYear{2022}
\setcopyright{acmlicensed}\acmConference[WSDM'22]{Proceedings of the Fifteenth ACM International Conference on Web Search and Data Mining}{February 21--25, 2022}{Tempe, AZ, USA}
\acmBooktitle{Proceedings of the Fifteenth ACM International Conference on Web Search and Data Mining (WSDM'22), February 21--25, 2022, Tempe, AZ, USA}
\acmPrice{15.00}
\acmDOI{10.1145/3488560.3498527}
\acmISBN{978-1-4503-9132-0/22/02}


\maketitle

\input{intro}
\vspace{-0.12in}
\input{model}
\vspace{-0.05in}
\input{solution}
\vspace{-0.05in}
\input{eval}

\vspace{-0.05in}
\input{relate}
\vspace{-0.05in}
\input{conclusion}

\clearpage

\bibliographystyle{ACM-Reference-Format}
\bibliography{sigproc}


\end{document}

%% file: intro.tex
\section{Introduction}
\label{sec:intro}

Recommender systems have emerged as critical components to alleviate information overloading for users in various online applications, \eg, e-commerce~\cite{wang2018billion}, online video platform~\cite{wei2020graph} and social media~\cite{qi2021personalized}. The goal is to learn user preference and forecast the items that he or she will consume based on observed user behaviors.

Among various recommendation techniques, collaborative filtering (CF) has become the most promising recommendation architecture to model historical user interactions over items~\cite{zheng2018spectral,chen2017attentive}. Commonly, the core of existing CF paradigm is to project users and items into latent representation space such that their interaction structural information is preserved. For example, Autoencoder has been employed as the effective embedding function for the representation projection in AutoRec~\cite{sedhain2015autorec} and CDAE~\cite{wu2016collaborative}. To inject the high-order connection signals in CF, another promising research line model user-item interactions as a graph and generate the user/item feature representations with the graph structural information preserved. These models perform the message passing over the interaction graph to generate node-level embeddings layer by layer, such as PinSage~\cite{ying2018graph}, NGCF~\cite{wang2019neural} and LightGCN~\cite{he2020lightgcn}.

However, the majority of existing recommendation models assume that only a single type of interaction exists between user and item, whereas in practical recommendation scenarios are multiplex in nature~\cite{gao2019learning,wang2020beyond}. Taking the online retail platform as an example, users can interact with items in multiple manners, including page view, add-to-favourite and purchase. Different types of behaviors may characterize user preference from different intention dimensions and complement with each other for better user preference learning~\cite{tanjim2020attentive}. Therefore, it is challenging but valuable to capture behavior multiplicity and the underlying dependencies in recommendation. To address this challenge, existing work models the behavior dependency by introducing different aggregation schemes to integrate type-specific behavior embeddings, to enhance the representation on target user behaviors (\eg, customer purchase)~\cite{xia2020multiplex,jin2020multi,xia2021knowledge}. For example, MATN~\cite{xia2020multiplex} adopts the self-attention to encode the pairwise correlations between different types of behaviors, and make predictions on the target behaviors. A relation-aware embedding propagation layer is developed to learn the behavior multiplicity in MBGCN~\cite{jin2020multi}, to gather multi-behavior interaction information from high-order neighbors.

Despite the effectiveness of existing methods, these studies share two common limitations: \emph{First}, Sparse Supervision Signal under Target Behaviors: the most of current multi-behavior recommender systems are trained with supervised information in an end-to-end manner. That is to say, for making forecasting on the target user behaviors, it is required to have sufficient labeled data corresponding to the target behaviors (\eg, user purchase data). Unfortunately, the observed interactions under the target behavior type, are often sparse as compared with other types of user-item interactions. For example, purchase prediction task in online retail system still faces the challenge of lacking of ground-truth labels~\cite{2019online}. Hence, directly integrating type-specific behavior embeddings will sacrifice the performance due to lacking supervision signals of target behaviors. \emph{Second}, Personalized Multi-Behavior Patterns: multi-behaviour patterns may vary by users. Semantics of multi-typed user-item interactions and their mutual relationships are diverse, depending on the personalized characteristics of users~\cite{lu2020meta}. Without considering diverse user intents which motivates different types of user behaviors, previous modeling of multiplex user-item relationships leads to suboptimal representations.\\\vspace{-0.1in}

\noindent \textbf{Contributions}. Having realized the above challenges for recommendation with behavior multiplicity, we focus on exploring diverse multi-behavior patterns under a contrastive self-supervised learning prototype. Towards this end, this work proposes a new model-\textbf{\underline{C}}ontrastive \textbf{\underline{M}}eta \textbf{\underline{L}}earning (\model) for multi-behavior recommendation. In \model, we design a multi-behavior contrastive learning framework to capture the cross-type interaction dependency from different behavior views. This endows our developed recommender system to effectively distill additional supervision signal from different types of user behaviors, which augments the model optimization process with sparse supervision labels. Inspired by the recent success achieved by self-supervised representation learning, we leverage the idea of contrastive learning to design cross-type behavior dependency modeling task with the user self-discrimination. The goal of our multi-behavior contrastive learning is to reach the agreement between user's type-specific behavior representations via the constructed contrastive loss. In addition, to handle the preference diversity of users and capture the personalized multi-behavior patterns, we design contrastive meta network to characterize the customized behavior heterogeneity, empowering \model\ to maintain dedicated representations for different users. Our meta contrastive encoder first extracts the personalized meta-knowledge from users, and then feeds it into our weighting function for customized multi-behavior dependency modeling.

In a nutshell, this work makes the following contributions:

\begin{itemize}[leftmargin=*]

\item We propose a new multi-behavior learning paradigm \model\ for recommendation by emphasizing the importance of diverse and multiplex user-item relationships, as well as tackling the label scarcity problem for target behaviors.\\\vspace{-0.1in}


\item In our \model\ framework, we design a multi-behavior contrastive learning paradigm to capture the transferable user-item relationships from multi-typed user behavior data, which incorporates auxiliary supervision signals into the sparse target behavior modeling. Furthermore, our proposed meta contrastive encoding scheme allows \model\ to preserve the personalized multi-behavior characteristics, so as to be reflective of the diverse behavior-aware user preference under a customized self-supervised framework.\\\vspace{-0.1in}


\item We perform extensive experiments on three real-world recommendation datasets to justify the rationality of our assumptions and the effectiveness of our proposed framework. By comparing \model\ with 12 baselines, we show that \model\ is able to consistently improve the performance of different techniques under various settings. Further analysis demonstrates the effectiveness of the designed sub-modules with ablation study.

\end{itemize}


%% file: model.tex
\section{Preliminary}
\label{sec:model}

We first define $\mathcal{U}$ and $\mathcal{I}$ to represent the set of users and items, respectively. In our multi-behavior recommendation scenario, let $\mathcal{X}^{(k)}$ denote the user-item interaction matrix under the $k$-th behavior type (\eg, page view, add-to-favorite, purchase). Hence, multi-behavior interaction data is represented as \{$\mathcal{X}^{(1)}$,..., $\mathcal{X}^{(k)}$,...,$\mathcal{X}^{(K)}$\}, where $K$ is the number of behavior types. In particular, the element $x_{u,i}^k=1$ indicates that user $u$ has interacted with item $i$ under the behavior type of $k$ before, and $x_{u,i}^k=0$ otherwise. Generally, there exist \emph{target behavior} as the prediction objective. Other types of user behaviors serve as the \emph{auxiliary behaviors}. For example, purchases are directly related to Gross Merchandise Value (GMV) in E-commerce services, and are usually considered as the target behaviors in various user modeling applications. Auxiliary behaviors could be the interactions of page view and add-to-favorite/cart.\\\vspace{-0.12in}

\noindent \textbf{Problem Statement}. The studied task is formally stated as: \textbf{Input}: observed user-item interactions with multiplex $K$ types of behaviors \{$\mathcal{X}^{(1)}$,..., $\mathcal{X}^{(k)}$,...,$\mathcal{X}^{(K)}$\} among users $\mathcal{U}$ and items $\mathcal{I}$. \textbf{Output}: a predictive function which estimates the likelihood of user $u$ will interact with item $i$ under the target type ($k$) of behaviors.\\\vspace{-0.12in}

\noindent \textbf{Multi-Behavior Interaction Graph}. Inspired by the representation paradigm of graph collaborative filtering methods~\cite{wang2019neural,wang2020multi}, we explore the user-item graph structure for our multi-behavior recommendation scenario. Specifically, given $K$ types of user-item interaction matrices \{$\mathcal{X}^{(1)}$,..., $\mathcal{X}^{(k)}$,...,$\mathcal{X}^{(K)}$\}, we generate the multi-behavior interaction graph, in which the set of nodes $\mathcal{V} = \mathcal{U} \cup \mathcal{I}$ involves the user and item set. We further define the set of multiplex edges $\mathcal{E}$ to represent observed interactions with $K$ types of behaviors. In $\mathcal{E}$, edge $e_{u,i}^k$ between $u$ and $i$ indicates that $x_{u,i}^k=1$.







%% file: solution.tex
\section{Methodology}
\label{sec:solution}

\begin{figure*}
    \centering
    \includegraphics[width=\textwidth]{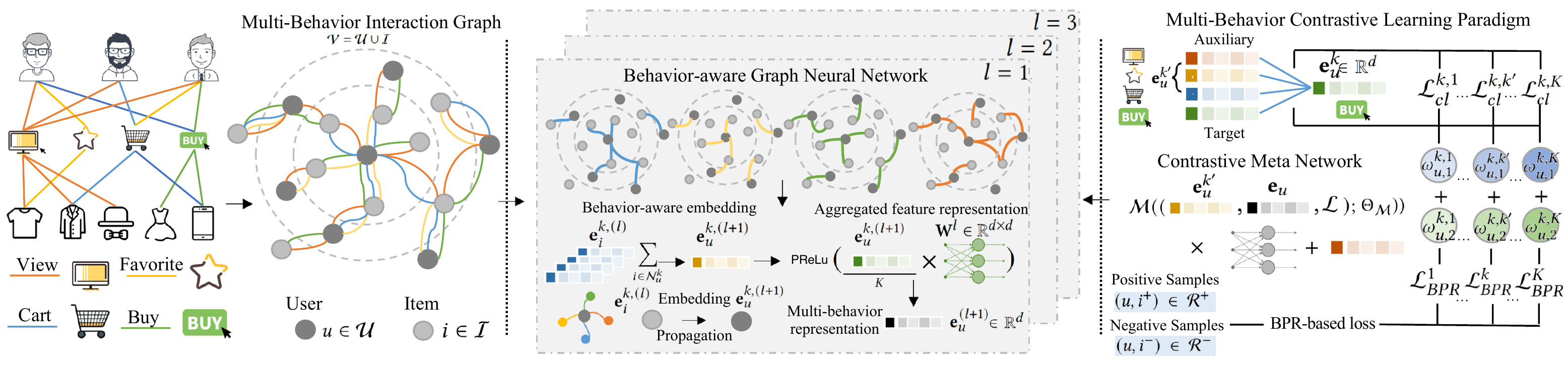}
    \vspace{-0.2in}
    \caption{The model flow of \model\ framework. i) The designed graph neural network $\mathcal{G}(\mathcal{A}; \Theta_{\mathcal{G}})$ performs the behavior-aware message passing over the multi-behavior interaction graph $G=\{\mathcal{V}, \mathcal{E}\}$. ii) The contrastive views are constrcuted between auxiliary and target behavior embeddings $\textbf{e}_u^k$, $\textbf{e}_u^{k'}$. iii) Our proposed meta contrastive encoder captures the customized cross-type behavior dependency with the meta weight network $\mathcal{M}((\mathcal{L}, \textbf{E}, \textbf{E}^k);\Theta_{\mathcal{M}})$. $\omega_{u}^{k,k'}$ is the personalized contrastive loss weight.}
    \vspace{-0.1in}
    \label{fig:framework}
\end{figure*}

We present our \full\ (\model) framework in this section, which encapsulates the customized meta learning into a self-supervised neural architecture, for personalized multi-behavior dependency modeling. The overall model flow is shown in Figure~\ref{fig:framework}. Key components will be elaborated in following subsections.

\subsection{Behavior-aware Graph Neural Network}
To inject the high-order connectivity into the multiplex relation learning across users/items, we first develop a graph-based message passing framework with the awareness of behavior context. Motivated by graph-based information propagation neural architecture~\cite{2019heterogeneous} and the findings in the state-of-the-art model LightGCN~\cite{he2020lightgcn,ji2021you}, our behavior-aware message passing scheme is built over a lightweight graph architecture, which can be represented:
\begin{align}
\textbf{e}_u^{k, (l+1)} = \sum_{i\in \mathcal{N}_u^k} \textbf{e}_i^{k, (l)};~~\textbf{e}_i^{k, (l+1)} = \sum_{u\in \mathcal{N}_i^k} \textbf{e}_u^{k, (l)}
\end{align}
\noindent where $\textbf{e}_v^{k, (l+1)}\in \mathbb{R}^d$ is defined as the obtained representation of node $v$ ($v\in\{u,i\}$) under the $l$-th graph neural layer. $\mathcal{N}_u^k$ and $\mathcal{N}_i^k$ denotes the neighboring nodes of item $i$ and user $u$, respectively. After encoding the behavior-specific interaction patterns of users, we propose to perform the embedding aggregation across different types of behaviour patterns with the following operation for user representations (similar aggregation is applied for item side):
\begin{align}
\textbf{e}_u^{(l+1)} = \text{PReLu} (\textbf{W}^{l} \cdot \frac{\sum_{k\in K} \textbf{e}_u^{k, (l+1)}}{K})
\end{align}
\noindent The aggregated feature representation $\textbf{e}_u^{(l+1)}$ could preserve multi-behavior contextual information. $\textbf{W}^{l} \in \mathbb{R}^{d\times d} $ represents the transformation matrix corresponding to $l$-th graph propagation layer.


\subsection{Multi-Behavior Contrastive learning}
In our \model\ framework, we propose a multi-behavior contrastive learning paradigm to capture the complex dependencies across different types of user interactions via a self-supervised principle. Conceptually, we utilize the idea of contrastive learning strategy for instance discrimination by contrasting positive and negative samples~\cite{you2020graph,qiu2020gcc}. Our contrastive learning architecture endows our main supervised task (\ie, target behavior prediction) with the auxiliary supervision signals from the auxiliary behaviors.

\subsubsection{\bf Contrastive View Generation}
In contrastive learning paradigm, it is important to generate appropriate views for constructing diverse representations for the method to contrast with~\cite{chen2020simple}. In our recommendation scenario with behavior multiplicity, we propose to consider each type of behaviors as individual view, which performs the contrastive learning between user embeddings in different behavior views. Different from current multi-behavior recommender systems (\eg, MATN~\cite{xia2020multiplex}, MBGCN~\cite{jin2020multi}) which merely rely on behavior-wise embedding combination for target behavior prediction, we conduct the data augmentation by incorporating auxiliary behavior contextual information as supervision signals. This design not only encodes the cross-type behavior dependency, but also alleviates the skewed data distribution across different types of user interaction data. 

\subsubsection{\bf Behavior-Wise Contrastive Learning Paradigm}
After establishing contrastive views from multi-behavior context, we further devise a behavior-wise contrastive learning paradigm between the target behaviors and auxiliary behaviors. In particular, different behavior views of the same user are considered as positive pairs, and the views of different users are sampled as negative pairs. Given the encoded target behavior representation $\textbf{e}_u^{k}$ from our graph neural architecture, the generated positive and negative pairs are $\{\textbf{e}_u^{k}, \textbf{e}_u^{k'} | u\in \mathcal{U} \}$ and $\{\textbf{e}_u^{k}, \textbf{e}_{u'}^{k'} | u, u'\in \mathcal{U}, u \neq u' \}$. The incorporated auxiliary supervision enables our model to still recognize user $u$ from different behavior views (\ie, $k$ and $k'$; $k,k' \in K$) and captures the latent relationships between the auxiliary behaviors and target behaviors. Meanwhile, for different users $u$ and $u'$, the contrastive loss aims to discriminate their behavior embeddings after data augmentation.

Following works~\cite{wu2021self,zhu2021graph}, we utilize the InfoNCE~\cite{oord2018representation} loss in our multi-view contrastive learning framework, to measure the distance between embeddings. We define our self-supervised learning loss with the objective of maximizing the Mutual Information (MI) between user representations through contrasting positive pairs with the sampled negative pair counterparts. The InfoNCE-based contrastive loss is calculated as below:
\begin{align}
\label{eq:infoNCE}
\mathcal{L}_{cl}^{k,k'} = \sum_{u\in \mathcal{U}} -log \frac{\text{exp} (\varphi (\textbf{e}_u^{k}, \textbf{e}_u^{k'})/ \tau)  }{\sum_{u'\in \mathcal{U}} \text{exp} (\varphi(\textbf{e}_u^{k}, \textbf{e}_{u'}^{k'}) / \tau)}
\end{align}
\noindent Here, we define $\varphi(\cdot)$ as the similarity function (\eg, inner-product or cosine similarity) between two embeddings. $\tau$ represents the temperature hyperparameter for the softmax function. To sum up, we perform the contrastive learning via maximizing the agreement between two behavior views based on the above defined contrastive loss, and enforcing the divergence among different users. We obtain the contrastive loss $\mathcal{L}_{cl}^{k,k'}$ for each pair of target behavior ($k$) and auxiliary behavior ($k'$). Therefore, we generate a list of contrastive loss functions as: $\mathcal{L}_{cl} = \mathcal{L}_{cl}^{k,1}$ + ... + $\mathcal{L}_{cl}^{k,k'}$ + ... + $\mathcal{L}_{cl}^{k,K}$. 


\subsection{Meta Contrastive Encoding}
In our recommendation scenario, different users have various behaviour patterns and item interaction preferences. For example, some users are likely to pick up most of products from their favorite item list to purchase, while others may only buy sporadic products given that they add a lot of items with less interest into their list~\cite{lu2020meta}. The diversity of multi-behavior patterns from different users, results in different item interactions. Hence, effectively modeling the personalized dependencies across different types of behaviors, is also important in making accurate recommendations. To achieve this goal, we propose a meta contrastive encoding scheme to learn an explicit weighting function for the integration of multi-behavior contrastive loss. This module customizes our self-supervised learning paradigm with the diverse constrastive loss integration. Our meta contrastive encoding schema is a two-phase learning paradigm: i) We propose a meta-knowledge encoder to capture the personalized multi-behavior characteristics, so as to reflect the diverse behavior-aware user preferences. ii) Then, the extracted meta-knowledge will be incorporated into our developed meta weight network, to generate customized contrastive loss weight for cross-type behavior dependency modeling.

\subsubsection{\bf Meta-Knowledge Encoder}
In our meta contrastive encoding framework, we firstly extract the meta-knowledge to preserve user-specific behavior dependencies. Inspired by feature interaction mechanisms in~\cite{he2017neural,zhang2014start}, we design two types of meta-knowledge encoder with different integration techniques based on learned user behavior representations: $\textbf{e}_u$ and $\textbf{e}_u^{k'}$ (auxiliary behavior of $k'$):
\begin{align}
\textbf{Z}_{u,1}^{k,k'} = (d(\mathcal{L}_{cl}^{k,k'}) \cdot \gamma ) \parallel \textbf{e}_u^{k'} \parallel \textbf{e}_u;~~ \textbf{Z}_{u,2}^{k,k'} = \mathcal{L}_{cl}^{k,k'} \cdot ( \textbf{e}_u^{k'} \parallel \textbf{e}_u )
\end{align}
\noindent where the encoded meta-knowledge is represented by $\textbf{Z}_{u,1}^{k,k'}$ and $\textbf{Z}_{u,2}^{k,k'}$. We define $d(\cdot)$ as the duplicate function to generate a value vector corresponding to the embedding dimensionality. $\parallel$ denotes the concatenation operation. $\gamma$ is a scale factor for the enlarge value. With this design for learning the personalized characteristics, both the auxiliary-target behavior dependency and user-specific interaction context are preserved in the extracted meta-knowledge.

\subsubsection{\bf Meta Weight Network}
After encoding the meta knowledge with user-specific multi-behavior patterns, we design a weighting function $\xi(\cdot)$ mapping from meta-knowledge to contrastive loss weights. This module endows our recommendation framework with the capability of learning the multi-behavior relationships in a customized manner, to be reflective of personalized user preference under various types of behavior intentions. Formally, we define our weighting function as the following transformation layer:
\begin{align}
\xi(\textbf{Z}_{u}^{k,k'}) = \text{PReLU}(\textbf{Z}_{u}^{k,k'} \cdot \textbf{W}_{\xi} + \textbf{b}_{\xi} )
\end{align}
\noindent where $\textbf{W}_{\xi} \in \mathbb{R}^{d\times d}$ and $\textbf{b}_{\xi} \in \mathbb{R}^{d}$ represent the projection layer and bias term, respectively. Here, we utilize the PReLU activation function to incorporate non-linearity. On the basis of our meta weight network, we can obtain our personalized contrastive loss weight as follows:
\begin{align}
\omega_{u}^{k,k'} = \omega_{u,1}^{k,k'} + \omega_{u,2}^{k,k'} = \xi(\textbf{Z}_{u,1}^{k,k'}) + \xi(\textbf{Z}_{u,2}^{k,k'})
\end{align}
\noindent For each user $u$, $\omega_{u}^{k,k'}$ weight represents the customized explicit dependence between the target behavior type of $k$ and auxiliary behavior type of $k'$. Accordingly, with our meta contrastive encoding scheme, we can generate two lists of loss weights for InfoNCE-based self-supervised loss and Bayesian personalized ranking (BPR)-based recommendation objective loss.

\subsection{The Learning Process of \model\ Framework}
In this section, we first introduce our optimization objective and then present the training strategy for our \model\ framework. Finally, the analysis on the time complexity of our model is provided.

\subsubsection{\bf Optimization Objective}
In the model inference of \model, we leverage the Bayesian Personalized Ranking (BPR) loss to learn parameters, which encourages the probability estimation of user's observed interaction to be higher than his/her unobserved counterparts. Formally, the behavior-specific BPR loss is defined as:
\begin{align}
\mathcal{L}_{BPR}^{k} = \sum_{(u,i^+,i^-)\in O_k} - \text{In} (\text{sigmoid} (\hat{x}_{u,i^+}^k - \hat{x}_{u,i^-}^k)) + \lambda || \Theta ||^2 
\end{align}
\noindent $O_k$ represents the pairwise training samples of $k$-th behavior type, i.e., $O_k =\{ (u, i^+, i^-) | (u, i^+) \in \mathcal{R}^+, (u, i^-) \in \mathcal{R}^- \}$. Here, $\mathcal{R}^+$ and $\mathcal{R}^-$ denotes the corresponding observed and unobserved interaction of user $u$. $\Theta$ represents the learnable parameters and the $L_2$ regularization is applied for alleviating overfitting issue.

\subsubsection{\bf Model Training}
In this work, we follow the training strategy of meta-learning methods in previous work~\cite{shu2019meta,franceschi2018bilevel}, by updating the parameters of our graph neural architecture (represented as $\mathcal{G}(\mathcal{A}; \Theta_{\mathcal{G}})$) and multi-behavior contrastive meta network (represented as $\mathcal{M}((\mathcal{L}, \textbf{E}, \textbf{E}^k);\Theta_{\mathcal{M}})$) in an alternative way. Here, $\mathcal{A}$ denotes the input adjacent matrix of behavior-aware user-item interaction graph. $\textbf{E}$ and $\textbf{E}^k$ represents the learned cross-type and behavior-specific embedding matrix of all users, respectively. The model training consists of three phases in an optimization loop to improve the training efficiency of our models. In particular: i) In the first stage, we integrate the behavior-aware graph neural network (with cloned state) and contrastive meta network, to learn initial parameter space of our multi-behavior contrastive encoder over the entire training data. ii) In the second stage, we refine the model parameters $\Theta_{\mathcal{M}}$ of our contrastive meta network based on the meta data. iii) After generating the personalized contrastive loss weights, we leverage the updated $\Theta_{M}$ to ameliorate the parameter $\Theta_{\mathcal{G}}$ of our graph neural network. We formally present the nested optimization process as follows ($B$ denote the size of training batch):
\begin{align}
\Theta_{\mathcal{G}}^* & = \mathop{\arg\min}_{\theta} \triangleq \sum_{k=1}^K \sum_{b=1}^B \Big ( \mathcal{M}((\mathcal{L}_{cl,k}^{train \cup meta}, \textbf{E}, \textbf{E}^k);\Theta_{\mathcal{M}}) \cdot \mathcal{L}_{cl,k}^{train} \nonumber\\
& + \mathcal{M}((\mathcal{L}_{bpr,k}^{train \cup meta}, \textbf{E}, \textbf{E}^k);\Theta_{\mathcal{M}}) \cdot \mathcal{L}_{bpr,k}^{train} \Big )
\end{align}


\subsubsection{\bf Model Complexity Analysis}
We analyze the complexity of our \model\ framework from several key components: i) the computational cost of our lightweight graph neural architecture is $O(L\times K\times  | \mathcal{R}^{k+}  |\times d)$ for performing message passing across graph layers. $|\mathcal{R}^{k+}|$ represents the number of non-zero elements in the adjacent matrix under the behavior of $k$, and $L$ denotes the number of information propagation layers. The operations of linear transformations and mean-pooling for multi-behavior aggregation takes $O(L\times (N+M)\times d\times (K+d))$ time. ii) Our meta contrastive encoder takes $O(K\times  | \mathcal{R}^{k+}  | \times d^{2})$ time overhead. iii) The cost of InfoNCE-based mutual information calculation is $O(B\times d)$ and $O(B\times S \times d)$ for the numerator and denominator (in Equation~\ref{eq:infoNCE}), respectively. Here, $S$ is the sampling size of contrastive learning for reducing the time complexity and increasing the randomness to achieve model robustness~\cite{wang2021self}. Therefore, our multi-behavior contrastive learning paradigm takes $O(K\times | \mathcal{R}^{k+} | \times S \times d)$ time per epoch. In conclusion, our model could achieve comparable time complexity with state-of-the-art multi-behavior recommendation techniques (\eg, MBGCN, EHCF).

%% file: eval.tex
\section{Evaluation}
\label{sec:eval}

To evaluate \model's performance, we conduct experiments on several real-world datasets by answering the following research questions:

\begin{table}[t]
    \caption{Statistics of experimented datasets}
\vspace{-0.1in}
    \label{tab:data}
    \centering
    \footnotesize
	\setlength{\tabcolsep}{0.6mm}
    \begin{tabular}{ccccc}
        \midrule
        Dataset&User \#&Item \#&Interaction \#& Interactive Behavior Type\\
        \hline
        Tmall& 31,882 & 31,232 & 1,451,219 & \{Page View, Favorite, Cart, Purchase\}\\
        IJCAI-Contest& 17,435 & 35,920 & 799,368 &\{Page View, Favorite, Cart, Purchase\}\\
        Retail Rocket& 2,174 & 30,113 & 97,381 &\{Page View, Cart, Transaction\}\\
        \hline
    \end{tabular}
\vspace{-0.2in}
\end{table}

\begin{table*}[t]
\caption{Performance comparison of all compared methods on different datasets in terms of \emph{NDCG}@10 and \emph{HR}@10}
\vspace{-0.1in}
\centering
\setlength{\tabcolsep}{0.9mm}
\begin{tabular}{|c|c|c|c|c|c|c|c|c|c|c|c|c|c|c|c|c|}
\hline
Dataset & Metric  & BPR & PinSage & NGCF  & LightGCN & SGL   & HGT   & HeCo  & NMTR  & MBGCN & MATN  & KHGT & EHCF & \textbf{\model}\ & Imprv. & \emph{p}-val.\\
\hline
\multirow{2}{*}{Tmall} & HR & 0.243 & 0.274 & 0.322 & 0.342 & 0.350 & 0.357 & 0.358 & 0.362 & 0.381 & 0.406 & 0.391 & \underline{0.433} & \textbf{0.543} & 25.4\% & $3e^{-5}$\\
\cline{2-17}
& NDCG & 0.143 & 0.151 & 0.184 & 0.205 & 0.210 & 0.210 & 0.199 & 0.215 & 0.213 & 0.225 & 0.232       & \underline{0.260} & \textbf{0.327} & 25.8\% & $2e^{-4}$\\
\hline
\multirow{2}{*}{\begin{tabular}[c]{@{}c@{}}IJCAI-\\ Contest\end{tabular}} & HR   & 0.163  & 0.176   & 0.256 & 0.257    & 0.249 & 0.250 & 0.262 & 0.294 & 0.304 & 0.369 & 0.317 & \underline{0.409} & \textbf{0.477} & 16.6\% & $9e{-5}$\\ 
\cline{2-17} 
& NDCG & 0.085 & 0.091  & 0.124 & 0.122 & 0.123 & 0.119 & 0.121 & 0.161 & 0.160 & 0.209 & 0.182       & \underline{0.237} & \textbf{0.283} & 19.4\% & $6e{-3}$\\
\hline
\multirow{2}{*}{\begin{tabular}[c]{@{}c@{}}Retail\\ Rocket\end{tabular}}  & HR & 0.235  & 0.247   & 0.260 & 0.261    & 0.263 & 0.305 & 0.297 & 0.314 & 0.308 & 0.301 & \underline{0.324} & 0.321 & \textbf{0.356} & 9.9\% & $1e^{-3}$\\
\cline{2-17} 
& NDCG & 0.146 & 0.139 & 0.140 & 0.152 & 0.165 & 0.176 & 0.178 & 0.201 & 0.181 & 0.181 & 0.202       & \underline{0.207} & \textbf{0.222} & 7.3\% & $1e^{-2}$ \\
\hline
\end{tabular}
\label{tab:overall_performance}
\end{table*}

\begin{itemize}[leftmargin=*]

\item \textbf{RQ1}: How effective is the developed \model\ framework to tackle the behavior multiplicity in recommendation?\\\vspace{-0.1in}

\item \textbf{RQ2}: How do different modules contribute to the performance of \model, such as the multi-behavior contrastive learning paradigm and meta contrastive encoder?\\\vspace{-0.1in}

\item \textbf{RQ3}: How does \model\ perform to alleviate interaction data sparsity, when competing with state-of-the-art methods?\\\vspace{-0.1in}

\item \textbf{RQ4}: How do different hyperparameter settings affect \model?\\\vspace{-0.1in}

\item \textbf{RQ5}: How is the model interpretation ability of our \model?

\end{itemize}


\subsection{Experimental Settings}


\subsubsection{\bf Datasets.} We evaluate the effectiveness of our proposed \model\ on three publicly available recommendation datasets. We present the statistical information in Table~\ref{tab:data}. \textbf{Tmall}: This dataset is collected from Tmall site--one of the largest E-commerce platform in China. The user behavior data contains various interactions: \emph{Page View}, \emph{Add-to-Favorite}, \emph{Add-to-Cart} and \emph{Purchase}. Following the setting in~\cite{xia2020multiplex}, we keep users with at least three purchases for training and test. \textbf{IJCAI-Contest}: This data was adopted in IJCAI15 Challenge from a business-to-customer retail system. It shares the same behavior types with the Tmall data, which are reflective of various user intention over items. \textbf{Retailrocket}: It is another benchmark dataset collected from Retailrocket recommender system. In this dataset, user interactions are consisted of \emph{Page View}, \emph{Add-to-Cart} and \emph{Transaction}. Following previous works for recommendation with multi-behaviors~\cite{xia2020multiplex,jin2020multi}, purchase behaviors are set as the target behaviors and other types of interactions are considered as the auxiliary behaviors.


\subsubsection{\bf Baselines.} We compare our \model\ with the following state-of-the-art methods from two groups: \emph{Single-Behavior} and \emph{Multi-Behavior} recommender systems. These methods leverage various techniques to improve the recommendation performance:

\noindent \textbf{Single-Behavior Recommendation Methods}:
\begin{itemize}[leftmargin=*]

\item \textbf{BPR}~\cite{rendle2012bpr}: It is a widely adopted matrix factorization model with the optimization criterion of Bayesian personalized ranking.

\item \textbf{PinSage}~\cite{ying2018graph}: This method defines the importance-based neighboring nodes to perform the graph convolution. In PinSage, the message passing paths are constructed through the random walk.

\item \textbf{NGCF}~\cite{wang2019neural}: it is a representative graph neural framework which captures the collaborative effects in the embedding function of users based on the convolutional message passing scheme.

\item \textbf{LightGCN}~\cite{he2020lightgcn}: it simplifies the graph convolution network-based recommendation architecture by removing the feature transformation and nonlinear activation operations.

\item \textbf{SGL}~\cite{wu2021self}: this method performs the self-supervised learning over the user-item interaction graph with data augmentation from different views (\eg, node and edge dropout). The integrated auxiliary task is on the basis of node self-discrimination.

\end{itemize}

\noindent \textbf{Multi-Behavior Recommendation Models}:

\begin{itemize}[leftmargin=*]

\item \textbf{NMTR}~\cite{gao2019neural}: it combines the multi-task learning framework and neural collaborative filtering to investigate multi-typed user interaction behaviors based on the predefined cascading relationships.

\item \textbf{MATN}~\cite{xia2020multiplex}: it adopts the attention mechanism for multi-behavior recommendation. Specifically, it uses memory-enhanced self-attention to measure the influence between different behaviors. The number of memory units is tuned from the range of [2,8].

\item \textbf{MBGCN}~\cite{jin2020multi}: this approach is a GCN-based model by capturing the multi-behavioral patterns over the constructed user-item interaction graph. The high-order connectivity is considered during the information propagation.

\item \textbf{KHGT}~\cite{xia2021knowledge}: this approach leverages transformer to incorporate the temporal information into the multi-behavior modeling, and differentiates the behaviors with graph attention network.

\item \textbf{EHCF}~\cite{chen2020efficient}: it conducts the knowledge transfer among heterogeneous user feedback to correlate behavior dependency. A new loss is used for model optimization from the positive-only data.

\end{itemize}

\noindent We further compare our \model\ with two state-of-the-art heterogeneous graph neural networks, by applying them to capture the heterogeneous behavior relations in recommendation.

\begin{itemize}[leftmargin=*]

\item \textbf{HGT}~\cite{hu2020heterogeneous}: This graph transformer models heterogeneous relations in graphs. We adopt the heterogeneous message passing schema to encode the multiplex behaviors with dedicated representations.

\item \textbf{HeCo}~\cite{wang2021self}: It is a recently developed heterogeneous graph neural network based on the cross-view supervised learning architecture. We generate the meta-path relation from our multi-behavior interaction graph.

\end{itemize}

\subsubsection{\bf Hyperparameters and Metrics}
We implement our \model\ with PyTorch. The embedding initialization is performed with Xavier~\cite{glorot2010understanding} and the model is optimized by adopting the AdamW optimizer~\cite{loshchilov2017decoupled} and the Cyclical Learning Rate (CyclicLR) strategy~\cite{smith2017cyclical}. In specific, the base and max learning rate is searched from \{$0.6e^{-4}$, $1e^{-4}$, $1e^{-3}$\} and \{$0.6e^{-3}$, $1e^{-3}$, $2e^{-3}$, $5e^{-3}$\}, respectively. For all graph-based baselines, the number of graph-based message propagation layers is tuned from \{1,2,3,4\}. We apply the L2 regularization for the learned embeddings with the weight tuned from \{$1e^{-3}$, $5e^{-3}$, $1e^{-2}$\}. Additionally, to alleviate the overfitting issue, the dropout is used in our designed meta network.

We adopt the widely used \emph{leave-one-out} strategy by generating the test set from users' last interacted items under the target behavior type (\ie, purchase/transaction). Two representative evaluation metrics are used for performance comparison: NDCG (Normalized Discounted Cumulative Gain) and HR (Hit Ratio). We also run our \model\ model and the best-performed baseline method for 10 times to calculate p-values for significance analysis.

\subsection{Performance Comparison (RQ1)}
We present the detailed evaluation results of all methods on different datasets in Table~\ref{tab:overall_performance} where the results of our \model\ and the best performed baselines are highlighted with bold and underlined, respectively. Key observations are as follows:

\begin{itemize}[leftmargin=*]

\item \model\ consistently outperforms all types of baselines on three datasets. The p-values are much less than 0.05, which indicates statistically significant improvements between our method and baselines. We attribute the significant performance improvements to the following two reasons: 1) Through the meta contrastive network, \model\ captures the multi-behavior dependencies in a customized manner; 2) The designed contrastive learning paradigm incorporates auxiliary self-supervised signals from different types of behavior dimensions, which offers informative gradients to the graph-based collaborative filtering architecture. \\\vspace{-0.1in}

\item Multi-behavior recommendation approaches (\eg, MBGCN, EHCF, KHGT) yield better performance than single-behavior recommendation methods (\eg, NGCF, LightGCN, PinSage), which reveals the helpfulness of exploring multi-behavioral information into the user preference modeling. Among various multi-behavior recommendation models, EHCF is the best baseline in most cases. This observation indicates that incorporating the different behavior semantics with supervision labels is able to guide the model optimization. Additionally, different from the topology-based self-supervised method-SGL, our \model\ designs new contrastive learning paradigm to fit the multi-behavior recommendation. \\\vspace{-0.1in}

\item \model\ outperforms heterogeneous graph neural networks (\ie, HGT and HeCo) by a large margin in all cases, verifying that our designed meta contrastive network endows the heterogeneous collaborative filtering with the capability of effectively encoding the relation heterogeneity.


\end{itemize}


\subsection{Ablation and Effectiveness Analyses (RQ2)}
To shed light on the performance improvement, we further conduct the ablation study for our \model, to justify the rationality of the designed key components. Analysis details are summarized as:

\begin{itemize}[leftmargin=*]

\item \textbf{\emph{Effect of multi-behavior contrastive learning framework}}. We first aim to answer the question: is it beneficial to integrate behavior-wise dependency under a contrastive learning prototype for \model. Towards this end, we generate a model variant \model(w/o)-CLF by disabling the contrastive learning between the target and auxiliary user behaviors. Instead, we only rely on the behavior-aware graph neural network to capture the behavior relationships. We present the evaluation results in Table~\ref{tab:module_ablation} with the following key summaries: 1) \model\ always outperforms \model(w/o)-CLF. This suggests the effectiveness of our contrasive learning paradigm, by capturing the complex dependent relations across different types of behaviors. 2) This design also mitigates the effect of skewed data distribution in the multi-behavior data, and effectively transfers knowledge from different behavior views. \\\vspace{-0.1in}

\item \textbf{\emph{Effect of meta contrastive network}}. To investigate whether the meta contrastive network benefit the multi-behavior dependency modeling, we propose another variant \model(w/o)-MCN which only conducts the contrastive learning between type-specific behavior embeddings based on the estimated mutual information. In other words, cross-behavior contrastive loss functions are integrated with the BPR-based loss using the equal weights, \ie, without explicitly differentiating the influence degrees under the augmented self-supervised learning tasks. Clearly, \model\ obtains better performance than \model(w/o)-MCN. It suggests that by employing the meta contrastive network, we can automatically discriminate the influence between different target-auxiliary behavior pairs. The cross-view behavior dependency can mutually complement with each other. \\\vspace{-0.1in}

\item \textbf{\emph{Effect of meta knowledge encoder}}. To verify the impact of meta knowledge encoder in our contrastive learning framework, we do an ablation study (with variant \model(w/o)-MKE) by disabling the meta contrastive weight network $\mathcal{M}(\cdot)$. Instead, we use a weighted gating mechanism to aggregate the behavior-specific contrastive loss in a uniform manner. Removing the incorporation of our meta knowledge degrades the performance, suggesting the necessity of our customized contrastive learning for different types of target-auxiliary behavior dependency.

\end{itemize}


\subsection{Model Performance on Alleviating Interaction Data Sparsity (RQ3)}
\begin{figure}[t]
	\centering
	\subfigure[][HR@10]{
		\centering
		\includegraphics[width=0.475\columnwidth]{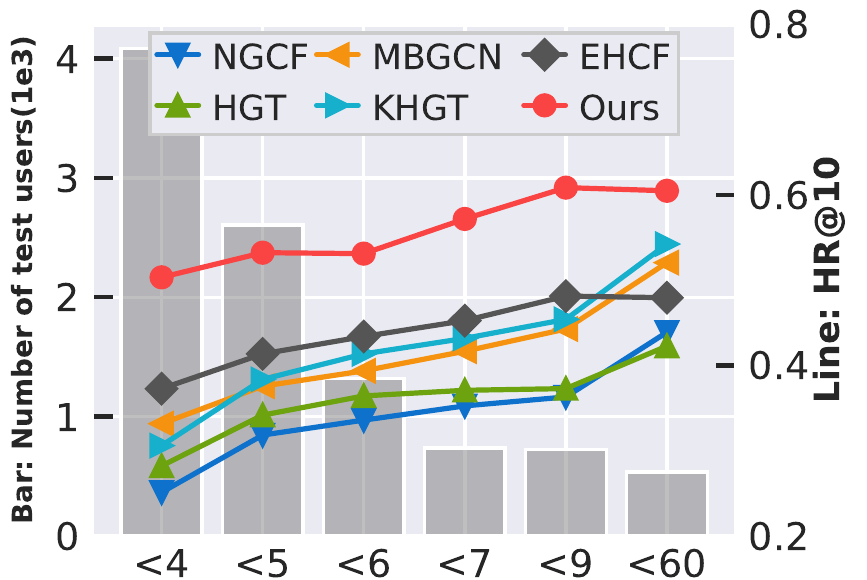}
		\label{fig:sp_tmall_hr}
	}
	\subfigure[][NDCG@10]{
		\centering
		\includegraphics[width=0.475\columnwidth]{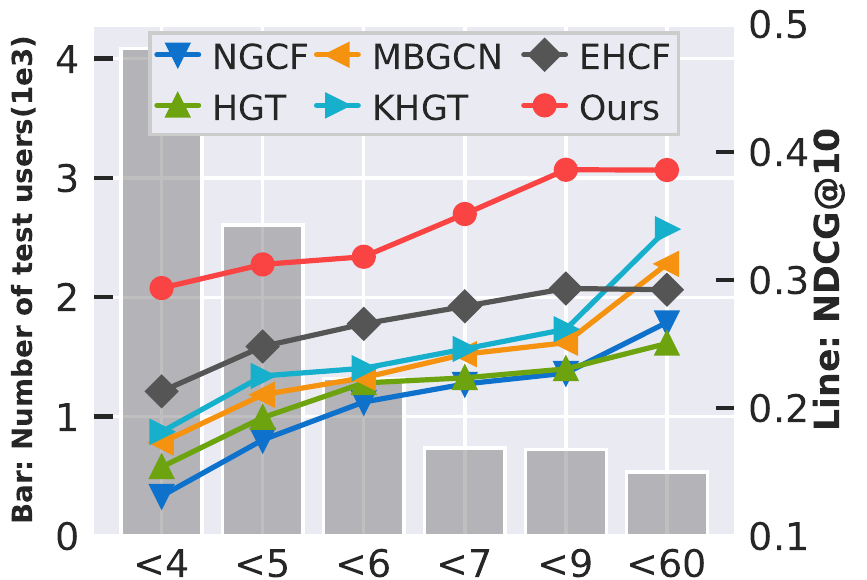}
		\label{fig:sp_tmall_ndcg}
	}
	\vspace{-0.15in}
	\caption{Performance comparison \wrt\ different interaction sparsity degrees on Tmall data.}
	\label{fig:sparsity}
	\vspace{-0.05in}
\end{figure}

In this section, we aim to show the rationality of bringing the contrastive learning into the multi-behavior recommendation, so as to alleviate the data sparsity issue. In Figure~\ref{fig:sparsity}, we show the evaluation result comparison with respect to different interaction sparsity degrees on Tmall data. Due to space limit, we select several representative baselines to make comparison. Specifically, we split users into six groups in terms of the number of their interactions (\eg, ``$\textless$7'' and ``$\textless$60''). The reported model performance measured by HR and NDCG (as shown in the right side of y-axis in Figure~\ref{fig:sparsity}) is averaged over all users in each group. The total number of users belonging to each group is shown in the left side of Figure~\ref{fig:sparsity}.



We have the following findings: i) The recommendation accuracy improves for all compared methods as the number of user interactions increases. It is reasonable since the quality behavior embeddings are more likely to be learned with sufficient user behaviors. ii) As compared to the vanilla collaborative filtering model (NGCF), multi-behavior recommender systems (\eg, KHGT, MBGCN) achieve better performance, suggesting the effectiveness of incorporating multi-typed behavior context for data sparsity alleviation. iii) \model\ consistently outperforms other multi-behavior recommendation methods under different interaction degrees. This observation indicates that \model\ solves the data sparsity issue better, by embracing the self-supervised contrastive learning paradigm for preserving the behavior heterogeneity in recommendation.

\begin{table}[t]
    \caption{Ablation study on key components of \model}
    \centering
    \small
    \begin{tabular}{c|cc|cc|cc}
        \hline
        Data & \multicolumn{2}{c|}{Tmall} & \multicolumn{2}{c|}{IJCAI-Contest} & \multicolumn{2}{c}{Retailrocket}\\
        \hline
        Metrics & HR & NDCG & HR & NDCG & HR & NDCG\\
        \hline
        \hline
        w/o-CLF  & 0.4665 & 0.2752 & 0.3636 & 0.1978 & 0.3032 & 0.1864 \\
        w/o-MCN & 0.5211 & 0.3097 & 0.4527 & 0.2703 & 0.3523 & 0.2185 \\
        w/o-MKE  & 0.5237 & 0.2988 & 0.4601 & 0.2715 & 0.3506 & 0.2079 \\
        \hline
        \emph{\model} & \textbf{0.5431} & \textbf{0.3266} & \textbf{0.4769} & \textbf{0.2829} & \textbf{0.3560} & \textbf{0.2219} \\
        \hline
    \end{tabular}
    \vspace{-0.15in}
    \label{tab:module_ablation}
\end{table}

\subsection{Hyperparameter Analysis on \model\ (RQ4)}

\begin{figure}[t]
    \centering
    \includegraphics[width=1\columnwidth]{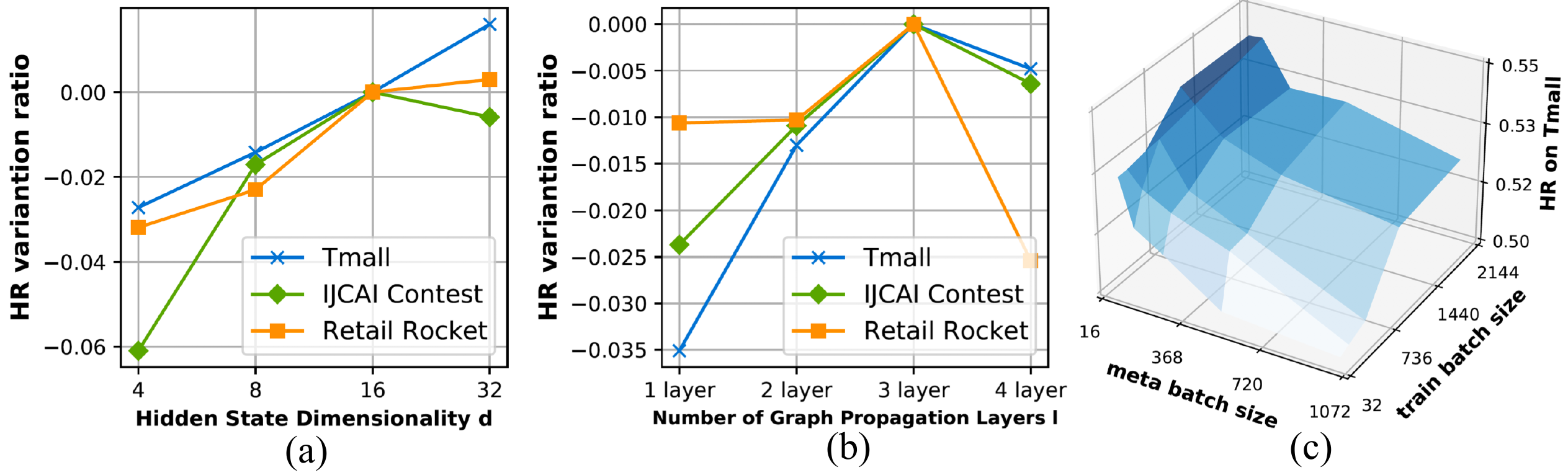}
    \vspace{-0.3in}
    \caption{Hyperparameter analysis of \model.}
    \vspace{-0.2in}
    \label{fig:parameter_study}
\end{figure}

This section examines the impact of different settings of several key hyperparameters in our proposed \model\ framework, including \# graph propagation layers $L$, representation dimensionality $d$, batch size in training process. Figure~\ref{fig:parameter_study} reports the evaluation results. For each time, we investigate the effect of one hyperparameter at a time and keep other parameters with their default settings. \\\vspace{-0.1in}

\noindent \textbf{\# graph propagation layers $L$}. From Figure~\ref{fig:parameter_study}, we can observe that more graph propagation layers results in better performance when $L \leq 3$. This suggests that more message passing layers will capture latent dependency from high-order neighbors. When further stacking more graph layers might introduce noise to the user representations, which leads to the oversmoothing issue~\cite{chen2020measuring,luo2021learning}. \\\vspace{-0.1in}

\noindent \textbf{Representation dimensionality $d$}. Our model can achieve good performance with the embedding dimensionality $16 \leq d \leq 32$. It indicates that our \model\ can boost the performance with small hidden state dimensionality, This can be attributed to effectively enhancing the user-item interaction learning with multiplex relationships. \\\vspace{-0.1in}

\noindent \textbf{Batch size in learning process}. We search the batch size for our meta contrastive network (meta batch) and the graph neural architecture (train batch) from the range of \{128, 256, 512, 1024, 2048\} and \{256, 512, 1024, 2048, 4096\}, respectively. Darker color signals better performance in Figure~\ref{fig:parameter_study} (c). When the sampled batch size of meta network is smaller than that of base graph network, the model performance becomes better. This configuration will improve the cooperation between our augmented self-supervised learning task and BPR-based ranking objective.\vspace{-0.1in}

\subsection{Qualitative Evaluation (RQ5)}
In this section, we perform the qualitative evaluation to show the model interpretation with the learned meta contrastive weights across different behavior types. We also visualize the projected behavior embeddings to have a better understanding of our achieved agreement between type-specific behavior embeddings.\\\vspace{-0.1in}

\begin{figure}
    \centering
    \includegraphics[width=0.95\columnwidth]{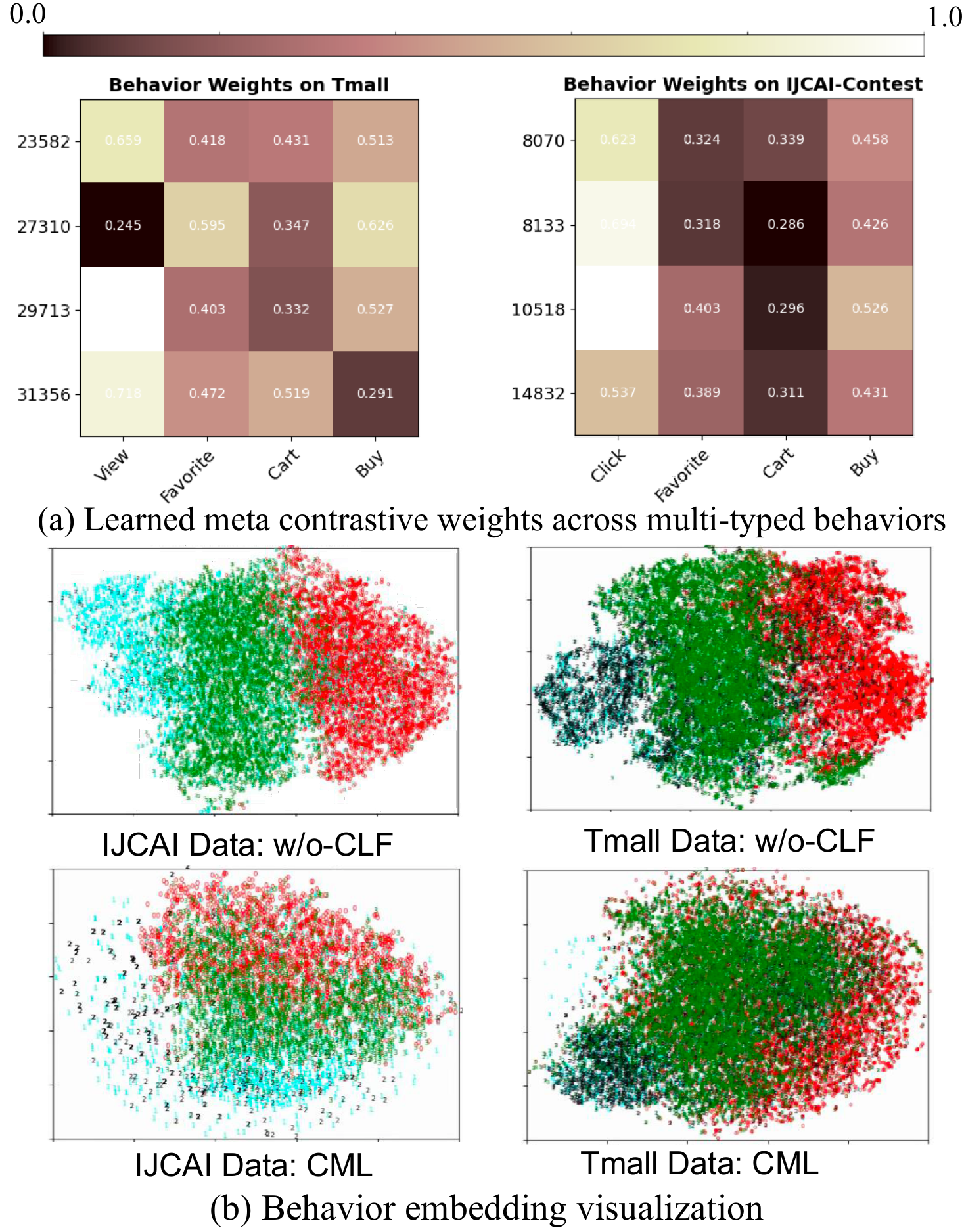}
    \vspace{-0.1in}
    \caption{Model interpretation study with (a) case studies of personalized contrastive weights from sampled different users; and (b) behavior embedding visualization, \ie, red: page view, blue: add-to-favorite, black: add-to-cart, green: purchase. Best viewed in colors.}
    \vspace{-0.25in}
    \label{fig:case_study}
\end{figure}

\noindent \textbf{Meta Contrastive Weight Visualization}. We visualize the learned meta contrastive weights $\omega_{u}^{k,k'}$ for each auxiliary behavior pairs ($k-k'$) from several sampled users. The customized contrastive weights can be observed in Figure~\ref{fig:case_study} (a), which reflect the personalized multi-behavior interaction patterns of different users. Each $\omega_{u}^{k,k'}$ value indicates the weight of individual contrastive loss between the target and auxiliary behavior views. For example, for user with id: 27310, the learned weights for the constructed view-buy and favorite-buy contrastive loss is 0.243 and 0.595, respectively. This suggests that this user is more likely to place the order after he or she adds the products into the favorite list, as compared with his/her page view behaviors. \\\vspace{-0.1in}

\noindent \textbf{Embedding Visualization}. We further show the visualization (2-D projection with t-SNE~\cite{van2008visualizing}) of user behavior embeddings encoded from \model\ and w/o-CLF on IJCAI-Contest data, respectively. In particular, we use different colors to represent different types of behaviors, \ie, red: page view, blue: add-to-favorite, black: add-to-cart, green: purchase. From Figure~\ref{fig:case_study} (b), we observe the embedding agreement achieved by our \model. This again justifies the effectiveness of our \model\ in alleviating data scarcity issue with the knowledge transfer across different types of behaviors, under our contrastive self-supervised learning architecture.

%% file: relate.tex
\section{Related Work}
\label{sec:relate}


\subsection{Graph-based Recommendation Models}
Recent studies have demonstrated the promising results offered by GNN-based recommendation models, by using different information propagation functions to aggregate embeddings over neighbors~\cite{he2020lightgcn,huang2021recent,fan2019metapath,huang2021graph,song2019session,chen2021graph}. For example, by stacking multiple embedding propagation layers, NGCF~\cite{wang2019neural} can gather information from neighboring nodes with high-order connectivity. To address the burdensome design of GCN-based message passing in NGCF, LightGCN~\cite{he2020lightgcn} omits the weight matrix and utilizes the sum-based pooling operation to obtain better recommendation performance. Additionally, to differentiate relations in recommendation, attention-based aggregation functions have been designed for fusing various information in recommender systems, such as social influence~\cite{2021knowledge,fan2019graph,song2019session}, knowledge graph embedding~\cite{liu2021contextualized,wang2019kgat}, textual information~\cite{wu2019reviews}. Specifically, GraphRec~\cite{fan2019graph} discriminates influence between users using graph-based attention mechanism. Wu~\etal~\cite{wu2019reviews} develops an attentional graph neural paradigm to enhance the user and item representations with textural information. Motivated by the above research works, our contrastive meta learning framework is built over the graph neural network to capture the behavior-aware collaborative effects between users and items.

\vspace{-0.1in}
\subsection{Multi-Behavior Recommender Systems}
Under the multi-typed user-item interactions, there exist some recent works attempting to designing effective approaches for handling behavior multiplicity~\cite{xia2020multiplex,jin2020multi,xia2021knowledge,chen2020efficient,xia2021graph}. In particular, the behavior-wise relationships are characterized by attention mechanism in~\cite{xia2020multiplex,xia2021knowledge}. MBGCN~\cite{jin2020multi} learns discriminative behavior representations using graph convolutional network. MATN~\cite{xia2020multiplex} considers the influences among different types of interactions with attentive weights for pattern aggregation. However, most of them are not designed with the sparse behavior data in mind. To fill this gap, we propose a new model with contrastive learning at behavior semantic levels, which provides auxiliary informative supervision signals for knowledge transferring between behavior types.

\vspace{-0.1in}
\subsection{Contrastive Representation Learning}
Self-supervised learning techniques have been demonstrated to be effective in learning representations from both image data~\cite{deng2020disentangled} and textual data~\cite{fu2021lrc}. It aims to learn quality discriminative representations by contrasting positive and negative samples from different views. For visual data, different data augmentation strategies (\eg, rotation~\cite{gidaris2018unsupervised}, color distortion~\cite{chen2020simple}) are used to generate negative instances. To better represent the graph topological structures, Deep Graph InfoMax (DGI)~\cite{velickovic2019deep} aims to maximize the mutual information between node embedding and graph representations based on the original and corrupted graphs. In addition, a model-agnostic recommendation model SGL~\cite{wu2021self} has been proposed to augment the supervised task of recommendation with auxiliary tasks. It performs dropout operations over the graph connection structures with different strategies, \ie, node dropout, edge dropout and random walk. SMIN~\cite{long2021social} is a social-aware recommendation method with generative self-supervision. Inspired by the existing contrastive learning paradigms, this work proposes a new graph contrastive representation framework with the adaptive multi-behavior modeling, by exploring various semantic aspects of user-item interactions.


%% file: conclusion.tex
\section{Conclusion}
\label{sec:conclusion}

In this paper, we develop a novel multi-behavior contrastive meta learning framework for recommendation. Our model learns user representations by preserving behavior heterogeneous context with the agreement between behaviors views constructed from our contrastive learning paradigm. The behavior-aware graph neural architecture with multi-behavior self-supervision bring benefits to the heterogeneous relational learning for recommendation. We perform comprehensive experiments using several real-world datasets to demonstrate the effectiveness of our proposed \model\ method, by comparing it with various state-of-the-arts.

In this paper, we take the initial step to capture the diverse multi-behavior patterns of users for recommendation under the self-supervised learning paradigm. In the future, it would be interesting to explore the pre-train model strategy of our \model\ for online user modeling applications (\eg, user profiling). Additionally, another meaningful future research direction can be extending our framework to learn disentangled representations of users, which could reflect the multi-dimensional user interests.
